\documentclass[12pt,subeqn,a4paper]{article}

\usepackage{amssymb,amsmath,amsfonts,amsthm, amscd, mathrsfs,helvet, mathtools}
\usepackage{bm, stmaryrd}
\usepackage{empheq, caption}
\usepackage[usenames, dvipsnames]{color}
\usepackage{graphicx,verbatim}
\usepackage{psfrag}
\usepackage[all]{xy}

\theoremstyle{plain}

\def\ha{\mbox{\small $\frac{1}{2}$}}

\def\C{\mathcal{C}}

\def\F{\mathcal{F}}

\def\S{\mathcal{S}}

\def\Loop{\mathcal{L}}

\DeclareMathOperator{\tr}{tr}

\newcommand{\gray}[0]{\color{Gray}}

\newcommand{\red}[0]{\color{Red}}
\definecolor{brightBlue}{rgb}{0,0,1}

\newcommand{\green}[0]{\color{Green}}
\definecolor{Violet}{rgb}{0.47,0,1}
\newcommand{\purple}[0]{\color{Violet}}

\newcommand{\tensor}[1]{{\bf \underline{#1}}}
\def\1{\tensor{1}}
\def\2{\tensor{2}}
\def\3{\tensor{3}}

\numberwithin{equation}{section}

\begin{document}

\begin{flushright}
DESY 11-078\\
{\bf May 2011}\\
\end{flushright}
\begin{centering}
\vspace{10mm} %
{\Large {\bf Splitting strings on integrable backgrounds}}\\

\vspace{15mm}
{\large Beno\^{\i}t Vicedo}\\
\vspace{4mm}
{\it DESY Theory, Notkestra\ss e 85, 22603 Hamburg, Germany}\\
\vspace{2mm}
\small{\tt benoit.vicedo@desy.de}

\vspace{10mm} 
{\bf Abstract} \\
\vspace{5mm} 
\end{centering}

We use integrability to construct the \emph{general} classical splitting string solution on $\mathbb{R} \times S^3$. Namely, given \emph{any} incoming string solution satisfying a necessary self-intersection property at some given instant in time, we use the integrability of the worldsheet $\sigma$-model to construct the pair of outgoing strings resulting from a split. The solution for each outgoing string is expressed recursively through a sequence of dressing transformations, the parameters of which are  determined by the solutions to Birkhoff factorization problems in an appropriate real form of the loop group of $SL_2(\mathbb{C})$.

\newpage

\input{epsf}

\section{Introduction}

Solving a conformal field theory amounts to finding its spectrum of anomalous dimensions and the structure constants of its 3-point functions.
At weak coupling, the former problem is equivalent to perturbatively diagonalizing the dilatation operator. When the CFT admits an AdS dual, the same problem at strong coupling requires finding the semiclassical energy spectrum of strings in AdS-space.
In the case of $\mathcal{N} = 4$ SYM, dual to type IIB superstrings on $AdS_5 \times S^5$, the emergence of integrability in the planar limit of both theories has gradually led to a very elegant analytical solution to both of these problems (see \cite{Beisert:2010jr} for a recent review).
Assuming integrability at all loops, this eventually culminated in a series of proposals for computing the full spectrum of anomalous dimensions at all values of the coupling \cite{GKV, Bombardelli:2009ns, AF}.

In sharp contrast with these developments for the spectrum of $\mathcal{N} = 4$ SYM, comparatively little is known about the structure constants of its 3-point functions. A possible explanation for this shortcoming might be the apparent lack of integrability methods beyond the planar sector. And yet it was recently shown in \cite{Tayloring1}, building on \cite{WeakIntegr}, that the integrability of planar $\mathcal{N} = 4$ SYM, which was so crucial in the exact study of the spectrum, can also be used to systematically tackle the problem of 3-point functions at weak coupling. At strong coupling, on the other hand, a similar use of the classical integrability of superstrings on $AdS_5 \times S^5$, which is after all a local property on the worldsheet, has not yet been exploited. Indeed, the semiclassical study of 3-point functions \cite{HeavyHeavyLight} has thus far been restricted to cases where the path integral is dominated by some finite-gap solution with cylindrical worldsheet. In particular, a comparison \cite{Tayloring2} of the results of \cite{Tayloring1} with string theory in the Frolov-Tseytlin limit could only be considered in the case where two of the three string states are ``heavy'', \emph{i.e.} semiclassical, while the third is ``light''. Nevertheless, an attractive proposal for computing 3-point functions at strong coupling using classical methods was put forward in \cite{Janik:2010gc}, which relies on finding classical Minkowskian solutions splitting/joining in $S^5$.

\begin{figure}[h]
\centering
\psfrag{X}{\scriptsize $X^{\mu}(\cdot, 0)$}
\psfrag{XI}{\scriptsize \red $X^{\mu}_{\rm I}(\cdot, 0)$}
\psfrag{XII}{\scriptsize \green $X^{\mu}_{\rm II}(\cdot, 0)$}
\psfrag{0}{\scriptsize $0$}
\psfrag{t}{\scriptsize $\tau$}
\includegraphics[height=35mm]{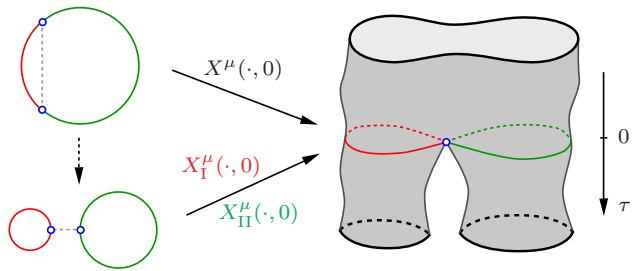}
\caption{The splitting is the result of a manual change in the topology of the closed string imposed at the instant $\tau = 0$ when the profile in space-time self-intersects. The full time evolution of the string is described by a worldsheet with the topology of a pair of pants.}
\label{fig: topology change}
\end{figure}

The aim of this paper is to exploit the classical integrability of the superstring $\sigma$-model on $AdS_5 \times S^5$ so as to construct classical string solutions with Lorentzian worldsheets of more general topology than the cylinder. The simplest such worldsheet is the `pair of pants' diagram, or three punctured sphere, describing the splitting or joining of strings. By focusing on the bosonic subspace $\mathbb{R} \times S^3$, we shall construct the most \emph{general} splitting string solution on this background. Specifically, we assume that we are given a solution on the cylinder with the property that: at some given instant in time $\tau = 0$, two of its points $\sigma_1$ and $\sigma_2$ coincide in target space and their velocities agree; a simple example is the folded spinning string \cite{Frolov:2003xy}. The splitting of such a string comes from treating it as consisting of two individual strings and letting each of them evolve separately, see Figure \ref{fig: topology change}. The problem therefore consists in solving a pair of Cauchy problems on the outgoing cylinders (the legs of the pair of pants) with Cauchy data specified by a portion of the original solution at time $\tau = 0$.

The classical splitting of strings has been extensively studied in the literature \cite{Splitting}, mostly in flat Minkowski space, but also on certain backgrounds with respect to which the equations of motion are linear in the fields. In each case, therefore, the general solution of the equations is given by a Fourier series which can be subjected to the relevant boundary conditions and Cauchy data in order to solve the Cauchy problem.
More recently, there has been renewed interest in splitting strings in the context of the AdS/CFT correspondence \cite{SplittingAdS}. However, in this case the non-linearity of the equations of motion renders Fourier analysis unapplicable.
But fortunately these non-linear equations are well known to be integrable, in the sense that they define a Lax connection: a locally defined 1-form on the worldsheet, meromorphic in some auxiliary complex parameter, and with the property of being flat. This will allow us to solve the Cauchy problems on the outgoing cylinders.

The outline of the paper is as follows. In section \ref{sec: setup} we set up the general formalism for discussing classical interacting strings, treating in parallel the case of flat space and $\mathbb{R} \times S^3$. In section \ref{sec: flat} we set up and solve the Cauchy problem for splitting strings in flat space. This serves as a warmup exercise since it will turn out that many features of the solution in flat space carry over to those of the Cauchy problem in $\mathbb{R} \times S^3$. Finally, section \ref{sec: S3} deals with splitting strings in $\mathbb{R} \times S^3$. After setting up the Cauchy problem on the outgoing cylinders, we show that the smoothness property of the solution is the same as in flat space, resulting in a similar tessellation of the worldsheet into tile-shaped regions bounded by null rays. We then show how integrability can be used to recursively construct the solution to the Cauchy problem in each tile, given the solution in the previous tile.

\section{Classical interacting strings} \label{sec: setup}

Our analysis of splitting strings on $\mathbb{R} \times S^3$ will be closely related to the corresponding analysis in flat space. In this section we therefore introduce both cases in parallel. Let $W$ denote the worldsheet of the string, equipped with a Lorentian metric $\gamma$. For the moment we impose no restriction on the topology of $W$.

\paragraph{In flat space.} Consider the embedding $X^{\mu} : W \to \mathbb{R}^{p,1}$ of $W$ into Minkowski space $\mathbb{R}^{p,1}$. In order to describe the classical motion of a string, this map should minimize the string action $\int_W dX^{\mu} \wedge \ast dX_{\mu}$, where $\ast$ denotes the Hodge dual relative to the worldsheet metric $\gamma$. The corresponding equations of motion for all the fields $(X^{\mu}, \gamma)$ read
\begin{subequations} \label{string eom flat}
\begin{align}
\label{string eom flat a} X^{\mu}: &\qquad d \ast d X^{\mu} = 0,\\
\label{string eom flat b} \gamma: &\qquad G^{\alpha \beta} = \ha \gamma^{\alpha \beta} \gamma_{\rho \sigma} G^{\rho \sigma},
\end{align}
\end{subequations}
where $G_{\alpha \beta} = \partial_{\alpha} X^{\mu} \partial_{\beta} X_{\mu}$ is the pull-back of the flat target space metric to $W$.

\paragraph{On $\mathbb{R} \times S^3$.} The embedding of a string with worldsheet $W$ into the target space $\mathbb{R} \times S^3$ with signature $(-1, +1, +1, +1)$ is described by a pair of fields $X_0 : W \rightarrow \mathbb{R}$ and $g : W \rightarrow SU(2)$. The action for all these fields can be written down succinctly as
\begin{equation} \label{string action}
S = \frac{\sqrt{\lambda}}{4 \pi} \int \left[ \frac{1}{2} \tr (j \wedge \ast j) + dX_0 \wedge \ast dX_0 \right],
\end{equation}
where $j = - g^{-1} dg$ is the $\mathfrak{su}(2)$-valued current and $\ast$ denotes the Hodge dual relative to the worldsheet metric $\gamma$. The resulting equations of motion for the respective fields are
\begin{subequations} \label{string eom}
\begin{align}
\label{string eom a} g: &\qquad d \ast j = 0,  \quad dj - j \wedge j = 0,\\
\label{string eom b} X_0: &\qquad d \ast d X_0 = 0,\\
\label{string eom c} \gamma: &\qquad G^{\alpha \beta} = \ha \gamma^{\alpha \beta} \gamma_{\rho \sigma} G^{\rho \sigma},
\end{align}
\end{subequations}
where the induced metric is $G_{\alpha \beta} = \ha \tr (j_{\alpha} j_{\beta}) + \partial_{\alpha} X_0 \partial_{\beta} X_0$.

\subsection{Mandelstam diagrams}

In the context of Riemannian worldsheets $W$, the conformal class of the Riemannian metric $g$ endows $W$ with a complex structure, promoting it to a Riemann surface $(W, [g])$. Moreover, as shown in \cite{GiddingsWolpert}, there is a 1 \---\ 1 correspondence between string light-cone diagrams and Abelian differentials on $W$.
Analogous statements to these can also be made in the Lorentzian setting \cite{Liu:1996daa}.
In particular, the conformal class of the Lorentzian metric $\gamma$ endows $W$ with a causal structure and static gauge provides a Mandelstam diagram representation of $W$.

\paragraph{Conformal gauge.} The equations of motion \eqref{string eom flat} or \eqref{string eom} being invariant under conformal transformations of the worldsheet metric $\gamma \mapsto e^{\phi} \gamma$, only the conformal equivalence class $[\gamma]$ of $\gamma$ is physically relevant. 
Yet there is a 1 \---\ 1 correspondence between conformal equivalence classes of Lorentzian metrics and causal structures on $W$, \emph{i.e.} ordered pairs $(\F_+, \F_-)$ of transverse null foliations \cite{Liu:1996daa}. In other words, specifying the worldsheet metric $\gamma$ amounts to giving $W$ a causal structure, thereby promoting it to a Lorentz surface $(W, [\gamma])$. In terms of any local null coordinates $\tilde{\sigma}^{\pm} : U \subset W \to \mathbb{R}$, defined by $\partial_{\pm} \coloneqq \partial_{\tilde{\sigma}^{\pm}}$ being tangent to the foliation $\F_{\mp}$, the metric reads $\gamma = \gamma_{+-} d\tilde{\sigma}^+ d\tilde{\sigma}^-$.

\paragraph{Static gauge.} We can write $dX_0 = \mu_+ + \mu_-$ where the pair $(\mu_+, \mu_-)$ are transverse measures to the foliations $(\F_+, \F_-)$ respectively, which locally read $\mu_{\pm} = (\partial_{\pm} X_0) d\tilde{\sigma}^{\pm}$. In other words $\mu_{\pm}$ vanishes on tangent vectors to leaves of the foliation $\F_{\pm}$ (in particular $\mu_{\pm}$ must vanish at singular points of the foliation $\F_{\pm}$ where multiple leaves end). By working in local coordinates it is easy to see that $\ast \mu_{\pm} = \pm \mu_{\pm}$. But now $dX_0$ is harmonic (that is, closed and co-closed) by the equations of motion, which means that $\mu_{\pm}$ are both closed and hence locally read $\mu_{\pm} = f_{\pm}(\tilde{\sigma}^{\pm}) d\tilde{\sigma}^{\pm}$.
Now integrating gives
\begin{equation*}
X_0(p) = \int^p (\mu_+ + \mu_-)
\end{equation*}
for $p \in W$, and since $X_0$ must be a well defined function we have $\int_C \mu_+ = - \int_C \mu_-$ for any $C \in \pi_1(W)$.
This last condition is the requirement for rectifiability of $W$ into a Mandelstam diagram \cite{Liu:1996daa}, \emph{i.e.} for there to be a representative of $(W, [\gamma])$ which is a Mandelstam diagram. Indeed, the level curves of $X_0$ are transverse to the leaves of both null foliations $\F_{\pm}$ and those through the singularities of $\F_{\pm}$ provide an annuli decomposition of $W$.
The representative of the class $[\gamma]$ is given by $\eta \coloneqq -4 \, \mu_+ \cdot \mu_- = -4\, d\sigma^+ d\sigma^- = - d\tau^2 + d\sigma^2$, where
\begin{equation} \label{global null coords}
\sigma^{\pm} \coloneqq \int^p \mu_{\pm} = \ha (\tau \pm \sigma)
\end{equation}
define global null coordinates on (the universal cover of) $W$. In particular we have $X_0 = \tau$.
Note that $\eta$ is degenerate at the singular points of the foliation $\F_{\pm}$.

\paragraph{Virasoro constraints.} After going to conformal static gauge, the dynamical equation for $X_0$ is solved and the metric is now fixed to the flat metric $\gamma = \eta$. The equations of motion then reduce to
\begin{align}
\label{string eom conf flat}
\textbf{In flat space} \quad &\left\{ \begin{array}{l} \partial_+ \partial_- X^{\mu} = 0, \quad \mu \neq 0,\\
T_{\pm\pm} \coloneqq \partial_{\pm} X^{\mu} \partial_{\pm} X_{\mu} = 0 \end{array} \right.
\\
\label{string eom conf S3}
\textbf{On $\mathbb{R} \times S^3$} \quad &\left\{ \begin{array}{l} \partial_- j_+ = \ha [j_-, j_+],  \quad \partial_+ j_- = \ha [j_+, j_-], \\
T_{\pm \pm} \coloneqq \ha \text{tr}\, j_{\pm}^2 + 1 = 0. \end{array} \right.
\end{align}
In either case, we find using the first set of equations that $T_{\pm\pm} = T_{\pm\pm}(\sigma^{\pm})$ only depends on $\sigma^{\pm}$. Therefore, if it vanishes for all $\sigma$ at some $\tau$ then it must vanish for all $\tau$.
It follows that when solving the Cauchy problem for the equations \eqref{string eom conf flat} or \eqref{string eom conf S3}, the Virasoro constraints $T_{\pm\pm} = 0$ will be automatically taken care of provided the Cauchy data satisfy them.

\subsection{Pair of pants}

\begin{figure}[h]
\centering
\psfrag{O}{\small O} \psfrag{I}{\small I} \psfrag{II}{\small II}
\includegraphics[height=30mm]{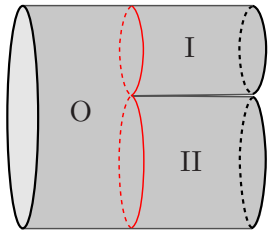}
\caption{The string worldsheet $W$ has the topology of a pair of pants, or three punctured sphere. The initial string O splits into two outgoing strings labelled I and II.}
\label{fig: pair of pants}
\end{figure}

Since our aim is to discuss splitting strings, from now on we shall focus on the case where $W$ has the topology of a pair of pants, see Figure \ref{fig: pair of pants}.
There is a single singular point and the level curve of $X_0$ through it, which has the topology of a `figure of 8', can be assumed to be at $\tau = 0$ without loss of generality. With this curve removed, the space $W \setminus \{\tau = 0\}$ consists of three cylinders, which can be parameterized as follows
\begin{subequations} \label{3 cyl}
\begin{align}
 \label{3 cyl O} W_{\rm O} &\coloneqq \{ (\sigma, \tau) \, | \, 0 < \sigma \leq 2 \pi, \, \tau < 0 \},\\
 \label{3 cyl I} W_{\rm I} &\coloneqq \{ (\sigma, \tau) \, | \, 0 < \sigma \leq 2 a \pi, \, \tau > 0 \},\\
 \label{3 cyl II} W_{\rm II} &\coloneqq \{ (\sigma, \tau) \, | \, 2 a \pi < \sigma \leq 2 \pi, \, \tau > 0 \}.
\end{align}
\end{subequations}
where $a < 1$ and the $\sigma$-interval is periodically identified in each case.

It is clear that any map from such a worldsheet $W$ into spacetime describes a single string which splits off into two separate strings at $\tau = 0$. In particular, the map has the following important self-intersection property:
\begin{align}
\label{self-intersecting flat}
\textbf{In flat space} \quad &\left\{ \begin{array}{c} \; X^{\mu}(0, 0) = X^{\mu}(2 a \pi, 0), \\
\partial_{\tau} X^{\mu}(0, 0) = \partial_{\tau} X^{\mu}(2 a \pi, 0) \end{array} \right.
\\
\label{self-intersecting S3}
\textbf{On $\mathbb{R} \times S^3$} \quad &\left\{ \begin{array}{c} g(0, 0) = g(2 a \pi, 0), \\
\partial_{\tau} g(0, 0) = \partial_{\tau} g(2 a \pi, 0). \end{array} \right.
\end{align}

\section{Splitting strings in flat space} \label{sec: flat}

Before studying splitting strings on $\mathbb{R} \times S^3$, we start by analyzing the splitting of strings in flat space in detail since many features of the solution will remain true in the nonlinear case and serve as a guideline there.

\subsection{Cauchy problem}

We would like to solve the Cauchy problem corresponding to the linear wave equation \eqref{string eom conf flat} on the pair of pants $W$, for a given set of initial conditions on the incoming circle. However, it is clear that these conditions cannot be completely arbitrary since they need to be such that the self-intersection property \eqref{self-intersecting flat} holds. We shall therefore assume that we are given a solution to \eqref{string eom conf flat} on the incoming cylinder $W_{\rm O}$ satisfying \eqref{self-intersecting flat}. To determine the complete solution on the rest of $W$, it remains to find solutions $X^{\mu}_{\rm I}$, $X^{\mu}_{\rm II}$ to
\begin{subequations} \label{eom I and II flat}
\renewcommand{\theequation}{\theparentequation{}.\Roman{equation}}
\begin{align}
\label{eom I flat} \partial_+ \partial_- X^{\mu}_{\rm I} &= 0, \qquad \text{on} \quad W_{\rm I} \\
\label{eom II flat} \partial_+ \partial_- X^{\mu}_{\rm II} &= 0, \qquad \text{on} \quad W_{\rm II} 
\end{align}
\end{subequations}
with initial conditions at $\tau = 0$ specified by the given solution $X^{\mu}$ on $W_{\rm O}$ as follows
\begin{subequations} \label{ic I and II flat}
\renewcommand{\theequation}{\theparentequation{}.\Roman{equation}}
\begin{alignat}{2}
\label{ic I flat} \left. \begin{array}{rcl} X^{\mu}_{\rm I}(\sigma, 0) \!\!&=&\!\! X^{\mu}(\sigma, 0),\\
\partial_{\tau} X^{\mu}_{\rm I}(\sigma, 0) \!\!&=&\!\! \partial_{\tau} X^{\mu}(\sigma, 0) \end{array} \right\} \quad &\text{for} &\quad 0 < &\;\sigma \leq 2 a \pi,\\
\label{ic II flat} \left. \begin{array}{rcl} X^{\mu}_{\rm II}(\sigma, 0) \!\!&=&\!\! X^{\mu}(\sigma, 0),\\
\partial_{\tau} X^{\mu}_{\rm II}(\sigma, 0) \!\!&=&\!\! \partial_{\tau} X^{\mu}(\sigma, 0) \end{array} \right\} \quad &\text{for} &\quad 2 a \pi < &\;\sigma \leq 2 \pi.
\end{alignat}
\end{subequations}
Note that for each outgoing string, the singularity lies at the end points of the initial interval. Since they live on $W_{\rm I}$ and $W_{\rm II}$ respectively, the solutions $X^{\mu}_{\rm I}$ and $X^{\mu}_{\rm II}$ defined for $\tau \geq 0$ are required to have new periodicity conditions, different from those of $X^{\mu}$, namely
\begin{subequations} \label{periodicity I and II flat}
\renewcommand{\theequation}{\theparentequation{}.\Roman{equation}}
\begin{align}
\label{periodicity I flat} X^{\mu}_{\rm I}(\sigma + 2 a \pi, \tau) &= X^{\mu}_{\rm I}(\sigma, \tau),\\
\label{periodicity II flat} X^{\mu}_{\rm II}(\sigma + 2 (1-a) \pi, \tau) &= X^{\mu}_{\rm II}(\sigma, \tau).
\end{align}
\end{subequations}
Notice that \eqref{periodicity I and II flat} is consistent with \eqref{ic I and II flat} by virtue of the self-intersection property \eqref{self-intersecting flat}.

\subsection{Absolute elsewhere of the singularity}

D'Alembert's solution to the linear wave equation is expressed as a sum of functions in $\sigma^+$ and $\sigma^-$. In particular, the solution given on the initial cylinder assumes this general form,
\begin{subequations} \label{sol flat}
\begin{equation} \label{sol initial}
X^{\mu}(\sigma, \tau) = X^{\mu}_+(\sigma^+) + X^{\mu}_-(\sigma^-).
\end{equation}
Since $X^{\mu}_{\text{I}}$, $X^{\mu}_{\text{II}}$ are solutions of the same equation \eqref{eom I and II flat}, they are also given by d'Alembert's general form on their respective cylinders
\begin{equation} \label{sol I and II flat}
X^{\mu}_{\rm I}(\sigma, \tau) = X^{\mu}_{{\rm I}+}(\sigma^+) + X^{\mu}_{{\rm I}-}(\sigma^-), \qquad
X^{\mu}_{\rm II}(\sigma, \tau) = X^{\mu}_{{\rm II}+}(\sigma^+) + X^{\mu}_{{\rm II}-}(\sigma^-).
\end{equation}
\end{subequations}
In terms of \eqref{sol flat} the initial conditions \eqref{ic I and II flat} read, in their respective domains in $\sigma$,
\begin{alignat*}{2}
X^{\mu}_{{\rm I}+}(\sigma) + X^{\mu}_{{\rm I}-}(\sigma) &= X^{\mu}_+(\sigma) + X^{\mu}_-(\sigma), &\quad \partial_{\sigma} X^{\mu}_{{\rm I}+}(\sigma) - \partial_{\sigma} X^{\mu}_{{\rm I}-}(\sigma) &= \partial_{\sigma} X^{\mu}_+(\sigma) - \partial_{\sigma} X^{\mu}_-(\sigma),\\
X^{\mu}_{{\rm II}+}(\sigma) + X^{\mu}_{{\rm II}-}(\sigma) &= X^{\mu}_+(\sigma) + X^{\mu}_-(\sigma), &\quad \partial_{\sigma} X^{\mu}_{{\rm II}+}(\sigma) - \partial_{\sigma} X^{\mu}_{{\rm II}-}(\sigma) &= \partial_{\sigma} X^{\mu}_+(\sigma) - \partial_{\sigma} X^{\mu}_-(\sigma).
\end{alignat*}
But then, differentiating the left set of equations with respect to $\sigma$ and combining them with the right set we obtain
\begin{alignat*}{3}
\partial_{\sigma} X^{\mu}_{{\rm I} \pm}(\sigma) &= \partial_{\sigma} X^{\mu}_{\pm}(\sigma), &\qquad &\text{for} &\quad 0 < &\;\sigma \leq 2 a \pi \\
\partial_{\sigma} X^{\mu}_{{\rm II} \pm}(\sigma) &= \partial_{\sigma} X^{\mu}_{\pm}(\sigma), &\qquad &\text{for} &\quad 2 a \pi < &\;\sigma \leq 2 \pi.
\end{alignat*}
Integrating in $\sigma$ and using part of the initial conditions again, we find
\begin{subequations} \label{sigma der flat}
\renewcommand{\theequation}{\theparentequation{}.\Roman{equation}}
\begin{alignat}{3}
X^{\mu}_{{\rm I} \pm}(\sigma) &= X^{\mu}_{\pm}(\sigma) \pm v^{\mu}_{\rm I}, &\qquad &\text{for} &\quad 0 < &\;\sigma \leq 2 a \pi\\
X^{\mu}_{{\rm II} \pm}(\sigma) &= X^{\mu}_{\pm}(\sigma) \pm v^{\mu}_{\rm II}, &\qquad &\text{for}  &\quad 2 a \pi < &\;\sigma \leq 2 \pi,
\end{alignat}
\end{subequations}
where $v^{\mu}_{\rm I}$ and $v^{\mu}_{\rm II}$ are constants. Plugging this into \eqref{sol I and II flat} we obtain
\begin{subequations} \label{sol diamond flat}
\renewcommand{\theequation}{\theparentequation{}.\Roman{equation}}
\begin{alignat}{3}
X^{\mu}_{\rm I}(\sigma, \tau) &= X^{\mu}(\sigma, \tau), &\qquad &\text{for} &\quad \tau < &\;\sigma \leq 2 a \pi - \tau,\\
X^{\mu}_{\rm II}(\sigma, \tau) &= X^{\mu}(\sigma, \tau), &\qquad &\text{for} &\quad 2 a \pi + \tau < &\;\sigma \leq 2 \pi - \tau.
\end{alignat}
\end{subequations}
Note that the domains of validity here are determined by the requirement that $\sigma^{\pm}$ satisfies the same bounds as $\sigma$ does in \eqref{sigma der flat}. Hence, these domains are bounded by null rays emanating from the singularity, as shown in Figure \ref{fig: regions I and II, R3}. That is, the original solution remains valid at points on the outgoing cylinders I and II which are space-like separated from the singularity.
\begin{figure}[h]
\centering
\begin{tabular}{ccc}
\psfrag{ap}{$2 a \pi$}
\psfrag{z}{$0$}
\psfrag{I}{I}
\psfrag{II}{II}
\psfrag{s}{$\sigma$}
\psfrag{t}{$\tau$}
\includegraphics[height=40.5mm]{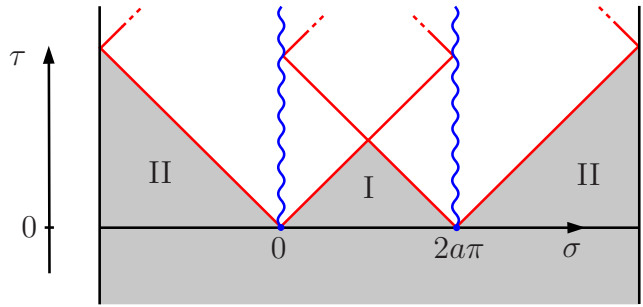}
\end{tabular}
\caption{When $\tau < 0$ the worldsheet has the topology of a cylinder, the picture being periodically identified in the $\sigma$-direction. For $\tau > 0$ the change in topology is indicated by the blue cuts emanating from the singularity at $0 \equiv 2 a \pi$: the right and left sides of the cut through $0$ are to be identified with the left and right sides of the cut through $2 a \pi$ respectively. In the shaded area, bounded by null rays (in red), the initial solution remains valid.}
\label{fig: regions I and II, R3}
\end{figure}

\subsection{Absolute future of the singularity}

To determine the solutions I and II beyond the limited region of Figure \ref{fig: regions I and II, R3}, we impose their respective periodicity conditions \eqref{periodicity I and II flat}.
Taking the derivatives of these conditions with respect to $\sigma^{\pm}$ and using the general form of the solutions \eqref{sol I and II flat} leads to
\begin{equation*}
\partial_{\sigma} X^{\mu}_{{\rm I}\pm}(\sigma + a \pi) = \partial_{\sigma} X^{\mu}_{{\rm I}\pm}(\sigma), \qquad
\partial_{\sigma} X^{\mu}_{{\rm II}\pm}(\sigma + (1 - a) \pi) = \partial_{\sigma} X^{\mu}_{{\rm II}\pm}(\sigma).
\end{equation*}
After integrating we find
\begin{equation*}
X^{\mu}_{{\rm I}\pm}(\sigma + a \pi) = X^{\mu}_{\text{I}\pm}(\sigma) + x^{\mu}_{{\rm I}\pm}, \qquad
X^{\mu}_{{\rm II}\pm}(\sigma + (1 - a) \pi) = X^{\mu}_{\text{II}\pm}(\sigma) + x^{\mu}_{{\rm II}\pm}.
\end{equation*}
In other words, the functions $X^{\mu}_{{\rm I}\pm}$ and $X^{\mu}_{{\rm II}\pm}$ are not periodic but shift by constants under the translations $\sigma \mapsto \sigma + a \pi$ and $\sigma \mapsto \sigma + (1-a) \pi$ respectively. This leads to
\begin{subequations} \label{sol extension flat}
\renewcommand{\theequation}{\theparentequation{}.\Roman{equation}}
\begin{gather}
\left\{
\begin{split}
X^{\mu}_{\rm I}(\sigma^+ + a \pi, \sigma^-) &= X^{\mu}_{\rm I}(\sigma^+, \sigma^-) + x^{\mu}_{\rm I},\\
X^{\mu}_{\rm I}(\sigma^+, \sigma^- + a \pi) &= X^{\mu}_{\rm I}(\sigma^+, \sigma^-) + x^{\mu}_{\rm I},
\end{split} \right. \\
\left\{
\begin{split}
X^{\mu}_{\rm II}(\sigma^+ + (1-a) \pi, \sigma^-) &= X^{\mu}_{\rm II}(\sigma^+, \sigma^-) + x^{\mu}_{\rm II},\\
X^{\mu}_{\rm II}(\sigma^+, \sigma^- + (1-a) \pi) &= X^{\mu}_{\rm II}(\sigma^+, \sigma^-) + x^{\mu}_{\rm II}.
\end{split} \right.
\end{gather}
\end{subequations}
That the \emph{same} $x^{\mu}_{\rm I}$ (resp. $x^{\mu}_{\rm II}$) appears for shifts in both $\sigma^+$ and $\sigma^-$ of $X^{\mu}_{\rm I}$ (resp. $X^{\mu}_{\rm II}$) follows from taking the difference of both equations and using the periodicity conditions \eqref{periodicity I and II flat} which can be written as $X^{\mu}_{\rm I}(\sigma^+ + a \pi, \sigma^-) = X^{\mu}_{\rm I}(\sigma^+, \sigma^- + a \pi)$ and similarly for $X^{\mu}_{\rm II}$.

The formulae \eqref{sol extension flat} now allow us to extend each solution $X^{\mu}_{\rm I}$, $X^{\mu}_{\rm II}$ beyond their restricted domains \eqref{sol diamond flat} depicted in Figure \ref{fig: regions I and II, R3 extend}. Together, equations \eqref{sol diamond flat} and \eqref{sol extension flat} therefore define the functions $X^{\mu}_{\rm I}$ and $X^{\mu}_{\rm II}$ completely on the whole outgoing cylinders $W_{\rm I}$ and $W_{\rm II}$. Combined with the original solution $X^{\mu}$ on the incoming cylinder, this gives a complete description of the corresponding splitting string.
Let us emphasize that since the construction assumed d'Alembert's form \eqref{sol I and II flat} for both functions $X^{\mu}_{\rm I}$ and $X^{\mu}_{\rm II}$, they automatically satisfy the equations of motion \eqref{eom I and II flat}, in a distributional sense, despite not being differentiable on the forward null rays through the singularity.

\begin{figure}[h]
\centering
\begin{tabular}{ccc}
\psfrag{pvI}{\footnotesize $+x_{\rm I}$}
\psfrag{pvII}{\footnotesize $+x_{\rm II}$}
\psfrag{p2vI}{\footnotesize $+ 2 x_{\rm I}$}
\psfrag{I}{I}
\psfrag{II}{II}
\psfrag{t}{$\tau$}
\psfrag{z}{$0$}
\includegraphics[height=51mm]{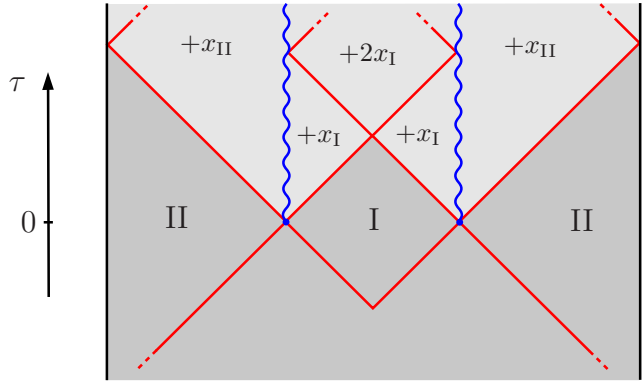}
\end{tabular}
\caption{The solution in the `null tiles' labelled I and II, delimited by null rays through the singularity, is obtained by extending the original solution $X^{\mu}$. The solutions $X^{\mu}_{\rm I}$ and $X^{\mu}_{\rm II}$ on subsequent tiles are given by constant translates of the solution in the regions I and II.}
\label{fig: regions I and II, R3 extend}
\end{figure}

One immediate advantage of this construction is that the qualitative description of the motion of the outgoing strings in space time is very transparent. In particular, the form of the solution clearly shows that the singular point, where the splitting occurs, propagates at the speed of light along the worldsheet of each outgoing string I and II. Along these null rays the space time profile of strings I and II exhibits cusps, but moreover, the profile away from these cusps is given by some rigid translate of a portion of the initial string.

\section{Splitting strings on $\mathbb{R} \times S^3$} \label{sec: S3}

\subsection{Cauchy problem}

As in the flat space case, to solve the Cauchy problem on the pair of pants $W$ we shall assume that a solution $g(\sigma, \tau)$ of the equations \eqref{string eom conf S3} on the incoming cylinder $W_{\rm O}$ is given, which satisfies the self-intersection property \eqref{self-intersecting S3} at $\tau = 0$. This solution will specify Cauchy data for separate Cauchy problems on each outgoing cylinder $W_{\rm I}$ and $W_{\rm II}$. However, since both problems are essentially equivalent we shall focus on one of the outgoing strings, say I.

The equations of motion for the embedding field $g_{\rm I} : W_{\rm I} \to SU(2)$ can be written as
\begin{subequations} \label{eom g system}
\begin{equation} \label{eom g}
\partial_{\tau}^2 g_{\rm I} - \partial_{\sigma}^2 g_{\rm I} = \partial_+ \partial_- g_{\rm I} = \ha \big( \partial_+ g_{\rm I} (g_{\rm I}^{-1} \partial_- g_{\rm I}) + \partial_- g_{\rm I} (g_{\rm I}^{-1} \partial_+ g_{\rm I}) \big) \eqqcolon f(g_{\rm I}, \partial_{\sigma} g_{\rm I}, \partial_{\tau} g_{\rm I}).
\end{equation}
Since the string is closed, we impose periodic boundary conditions, \emph{i.e.} $g_{\rm I}(0, \tau) = g_{\rm I}(2 a \pi, \tau)$. Equivalently we require $g_{\rm I}(\sigma, \tau)$ to be $2 a \pi$-periodic in $\sigma$. The initial conditions read
\begin{equation} \label{ic g S3}
g_{\rm I}(\sigma, 0) = g(\sigma, 0),\qquad 
\partial_{\tau} g_{\rm I}(\sigma, 0) = \partial_{\tau} g(\sigma, 0),
\end{equation}
\end{subequations}
where $g(\sigma, \tau)$ is the given initial string solution. By assumption it satisfies the self-intersection property \eqref{self-intersecting S3} so that \eqref{ic g S3} are consistent with the $2 a \pi$-periodicity of $g_{\rm I}$. Moreover, we assume the incoming string solution to be smooth so that \eqref{ic g S3} are both smooth except at the self-intersection point $\sigma = 0 \equiv 2 a \pi$ of the initial string, where they are only continuous. Such points of reduced regularity are referred to as `singularities' and the relation between singularities of a solution and singularities of the corresponding initial data goes under the name of `propagation of singularities'.

To solve the Cauchy problem \eqref{eom g system} we will proceed in three steps. First, we make use of the theory of propagation of singularities to identify the global smoothness properties of the outgoing string. It turns out that despite the nonlinearity of the equations, the singularity of the initial data propagates along null trajectories, exactly as in the flat space case where the equations were linear. We then argue that the initial solution can be trivially extended to all points which are spacelike separated from the singularity. Finally, using this information we construct the remainder of the solution in the forward light-cone of the singularity by exploiting the integrability of the equations and using the dressing method.

\subsection{Propagation of the singularity} \label{sec: prop sing S3}

The propagation of singularities in nonlinear Klein-Gordon type equations of the general form $\partial_{\tau}^2 u - \partial_{\sigma}^2 u = f(x, u, \partial_{\sigma} u, \partial_{\tau} u)$ was first studied in \cite{Reed} and further developed in \cite{ReedRauch1}. It turns out that despite the presence of nonlinear terms on the right hand side, the result is exactly the same as in the linear case where $f \equiv 0$. In other words, if the initial data is $C^n$ at $(\sigma, 0)$ then the solution will be $C^n$ with respect to $\partial_-$ along the right null ray $(\sigma + \tau, \tau)$ and $C^n$ with respect to $\partial_+$ along the left null ray $(\sigma - \tau, \tau)$. In particular, the solution will be smooth at a point $(\sigma, \tau)$ if its backward null rays intersect $\tau = 0$ only at non-singular points of the initial data. Note that the pair of null rays through any point are nothing but the characteristic lines of the second order hyperbolic differential operator $\partial_{\tau}^2 - \partial_{\sigma}^2$. This conclusion remains true more generally for coupled Klein Gordon equations such as \eqref{eom g} in which the nonlinear coupling terms depend only on lower order derivatives $g_{\rm I}$, $\partial_{\sigma} g_{\rm I}$, $\partial_{\tau} g_{\rm I}$.

The propagation of singularities in the light-cone components of the current $j_{\rm I} = - g_{\rm I}^{-1} dg_{\rm I}$ itself will be more relevant later so we discuss it directly. The equations of motion read
\begin{subequations} \label{eom j system}
\begin{equation} \label{eom jpm}
\partial_- j_{{\rm I}+} = \ha [j_{{\rm I}-}, j_{{\rm I}+}], \qquad \partial_+ j_{{\rm I}-} = \ha [j_{{\rm I}+}, j_{{\rm I}-}],
\end{equation}
and the corresponding initial conditions derived from \eqref{ic g S3} are
\begin{equation} \label{ic j S3}
j_{{\rm I} \pm}(\sigma, 0) = - g(\sigma, 0)^{-1} ( \partial_{\tau} g(\sigma, 0) \pm \partial_{\sigma} g(\sigma, 0) ).
\end{equation}
\end{subequations}
These functions are smooth away from the self-intersection point $\sigma = 0 \equiv 2 a \pi$, but exhibit a jump discontinuity there since $g(\sigma, 0)$ is only continuous at that point.

\begin{figure}[h]
\centering
\begin{tabular}{ccc}
\psfrag{Sp}{\scriptsize $\S_+$}
\psfrag{Sm}{\scriptsize $\S_-$}
\includegraphics[width=25mm]{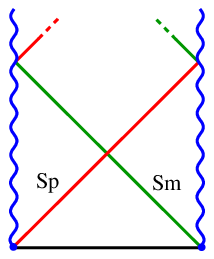} & $\qquad\qquad\qquad$ &
\psfrag{Sp}{\scriptsize $\S_+$}
\psfrag{Sm}{\scriptsize $\S_-$}
\psfrag{p}{\scriptsize $p$}
\includegraphics[width=25mm]{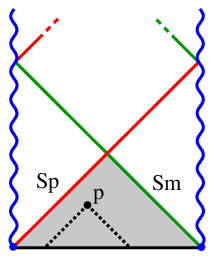}\\
$(a)$ &  & $(b)$
\end{tabular}
\caption{$(a)$ The self-intersection point of the initial string corresponds to a singularity in the initial conditions of each outgoing string, which propagates along both left and right null rays $\S_-$ and $\S_+$. The solution is therefore smooth everywhere except on these forward null rays. The component $j_{{\rm I} \mp}$ has a jump discontinuity across $\S_{\pm}$ whereas $j_{{\rm I} \pm}$ is continuous.\\
$(b)$ The shaded region, representing points which are causally disconnected from the singularity, is unaffected by the splitting. The solution there is simply given by the original solution extended beyond $\tau = 0$ as if the splitting never occurred.}
\label{fig: null characteristics}
\end{figure}

The general semilinear hyperbolic first order system with piecewise-smooth initial data having jump discontinuities only at a discrete set of points was studied in \cite{ReedRauch2}. 
The system \eqref{eom j system} is of this type but has only two characteristic directions at any point, namely the left and right null rays. It follows (see \cite{ReedRauch2} for details) that there can be no `anomalous' singularities \---\ these are singularities which are not present in the linearized system \---\ appearing at the intersection of two singularity bearing characteristics. Let us denote $\S_-$ and $\S_+$ the forward left and right null rays emanating from the singularity $(a \pi, 0)$, see Figure \ref{fig: null characteristics} $(a)$. Then a solution of \eqref{eom j system} in the distributional sense must in fact be smooth in $W_{\rm I} \setminus (\S_+ \cup \S_-)$.

We can be more precise about the value of the jump discontinuities across $\S_{\pm}$. By taking the integral of the first equation in \eqref{eom jpm} over a vanishingly small interval in the $\sigma^-$ direction which intersects $\S_+$, we find that the discontinuity of $j_{{\rm I}+}$ across this characteristic vanishes. That is, $j_{{\rm I}+}$ is continuous in the $\sigma^-$ direction. On the other hand, its discontinuity $\Delta^- j_{{\rm I}+}$ across the left null ray $\S_-$ satisfies $\partial_- (\Delta^- j_{{\rm I}+}) = \ha [j_{{\rm I}-}, \Delta^- j_{{\rm I}+}]$. A similar reasoning applied to the second equation in \eqref{eom jpm} shows that $j_{{\rm I}-}$ is continuous in the $\sigma^+$ direction and its jump discontinuity $\Delta^+ j_{{\rm I}-}$ across the right null ray $\S_+$ satisfies $\partial_+ (\Delta^+ j_{{\rm I}-}) = \ha [j_{{\rm I}+}, \Delta^+ j_{{\rm I}-}]$. The upshot is that the jump discontinuities of $j_{{\rm I}+}$ and $j_{{\rm I}-}$ in the initial condition at $\tau = 0$ propagate along the \emph{left} null ray $\S_-$ and \emph{right} null ray $\S_+$, respectively.

\subsection{Absolute elsewhere of the singularity} \label{sec: abs else S3}

Having identified the smoothness properties of the current $j_{\rm I}$ on each outgoing cylinder, we now proceed to actually construct these solutions. In the spirit of section \ref{sec: flat}, we will solve the equations successively in each `null-tile', delimited by the null lines $\S_{\pm}$, where the solution is known to be smooth. The first tile in contact with the Cauchy surface requires little effort.

Indeed, in any hyperbolic system, the solution of the Cauchy problem at any point $p$ only depends on that part of the Cauchy data which lies within the domain of dependence of $p$, defined as the interior of the backward characteristic cone with apex $p$. Now consider the region on the outgoing cylinder $W_{\rm I}$ consisting of all points, the domain of dependence of which does not contain the singularity. This region is delimited by the forward null rays $\S_{\pm}$ through the singularity and the Cauchy surface $\tau = 0$, see Figure \ref{fig: null characteristics} $(b)$. It is then clear that the original solution remains valid within this region, since the relevant Cauchy data is the same as if the splitting had never occurred.

\subsection{Integrability} \label{sec: integr}

Extending the solution into the region of influence of the singularity is considerably harder since it requires explicitly solving the Cauchy problem.
Fortunately, the equations \eqref{eom jpm} of the prinicpal chiral model are well known to be integrable in the sense that they can locally be rewritten in the form of a zero curvature equation. This will enable us to make use of powerful factorization methods to construct their solutions.

\paragraph{Lax connection.} Introduce the following one complex-parameter family of $\mathfrak{sl}_2(\mathbb{C})$-valued 1-forms on the worldsheet $W_{\rm I}$, depending on the single parameter $x \in \mathbb{C}P^1$,
\begin{equation} \label{Lax connection}
J_{\rm I}(x) \coloneqq \frac{j_{{\rm I}+}}{1 - x} \, d\sigma^+ + \frac{j_{{\rm I}-}}{1 + x} \, d\sigma^-,
\end{equation}
where recall that $j_{\rm I} = j_{{\rm I}+} d\sigma^+ + j_{{\rm I}-} d\sigma^-$. It has the remarkable property of being flat, \emph{i.e.}
\begin{equation} \label{flatness}
dJ_{\rm I}(x) - J_{\rm I}(x) \wedge J_{\rm I}(x) = 0,
\end{equation}
if and only if the equations of motion \eqref{eom jpm} hold. In other words, flat $\mathfrak{sl}_2(\mathbb{C})$-connections $J_{\rm I}(x)$ on $W_{\rm I}$ with simple poles at $x = \pm 1$ and a zero at $x = \infty$ are in 1 \---\ 1 correspondence with solutions $j_{\rm I} : W_{\rm I} \rightarrow \mathfrak{sl}_2(\mathbb{C})$ of the principal chiral model equations. Specifically, the Lax connection is constructed from $j_{\rm I}$ as in \eqref{Lax connection} and the current is recovered from $J_{\rm I}(x)$ by 
\begin{equation} \label{j from J}
j_{\rm I} = J_{\rm I}(0).
\end{equation}

\paragraph{Extended solution.} Since the $\mathfrak{sl}_2(\mathbb{C})$-connection $J_{\rm I}(x)$ is flat, it can be trivialized over any simply connected domain of $U \subset W_{\rm I}$, namely we can write
\begin{equation} \label{aux lin sys}
J_{\rm I}(x) = \big( d \Psi_{\rm I}(x) \big) \Psi_{\rm I}(x)^{-1},
\end{equation}
where $\Psi_{\rm I}$ is uniquely determined if we require $\Psi_{\rm I}(x, \sigma_0, \tau_0) = {\bf 1}$ at some point $(\sigma_0, \tau_0) \in U$. It follows from \eqref{aux lin sys} that $(d - \tr J_{\rm I}(x)) \det \Psi_{\rm I}(x) = 0$ and therefore $\det\Psi_{\rm I}(x)$ is constant since $\tr J_{\rm I}(x) = 0$. The initial condition then implies that $\Psi_{\rm I}(x)$ takes values in $SL_2(\mathbb{C})$.

Ultimately we are interested in the group element $g_{\rm I} \in SU_2$ rather than the current $j_{\rm I}$. Comparing the definition of $j_{\rm I} = -g_{\rm I}^{-1} dg_{\rm I}$ with that of $\Psi_{\rm I}(x)$ in \eqref{aux lin sys} and using \eqref{j from J}, we see that the group element can be recovered succinctly from $\Psi_{\rm I}(x)$ as
\begin{equation} \label{g from Psi}
g_{\rm I} = \Psi_{\rm I}(0)^{-1}.
\end{equation}
For this reason $\Psi_{\rm I}(x)$ is sometimes called the extended solution.

\paragraph{Gauge transformations.} The zero-curvature equation \eqref{flatness} has a large gauge redundancy since given any $\tilde{g}(x, \sigma, \tau)$, the gauge transformed Lax connection
\begin{equation} \label{gauge transf}
J_{\rm I}(x) \mapsto \tilde{g} J_{\rm I}(x) \tilde{g}^{-1} + (d \tilde{g}) \tilde{g}^{-1}
\end{equation}
also satisfies the zero-curvature equation. For generic choices of $\tilde{g}$, however, it will no longer admit the same pole structure as \eqref{Lax connection} and therefore can no longer be interpreted as a Lax connection of the principal chiral model. Yet, when $\tilde{g}$ is carefully chosen to preserve the pole structure of the Lax connection, \eqref{gauge transf} provides a powerful map between solutions.

\paragraph{Reality conditions.} To obtain $\mathfrak{su}_2$-valued currents $j_{\rm I}$ and $SU_2$-valued solutions $g_{\rm I}$, one must impose reality conditions on the extended solution $\Psi_{\rm I}(x)$. Sufficient conditions are
\begin{equation} \label{Psi reality}
\Psi_{\rm I}(x)^{\dag} = \Psi_{\rm I}(\bar{x})^{-1},
\end{equation}
which imply $g_{\rm I}^{\dag} = g_{\rm I}^{-1}$. Furthermore, the ensuing reality condition $J_{\rm I}(x)^{\dag} = - J_{\rm I}(\bar{x})$ on the Lax connection which follows from \eqref{aux lin sys} then implies $j_{\rm I}^{\dag} = - j_{\rm I}$. It will be convenient to think of real extended solutions \eqref{Psi reality} as fixed points of the complex antilinear involution
\begin{equation} \label{tau inv}
\hat{\tau}(\Psi_{\rm I})(x) \coloneqq \tau \big( \Psi_{\rm I}(\bar{x}) \big),
\end{equation}
where $\tau(A) \coloneqq (A^{\dag})^{-1}$ for any $A \in SL_2(\mathbb{C})$, so that $SU_2 \subset SL_2(\mathbb{C})$ is the fixed point set of $\tau$.

\subsection{Absolute future of the singularity}

When deriving exact solutions for the outgoing strings in the flat space case we made full use of the fact that d'Alembert's general solution to the linear wave equation is expressed as a linear superposition of two independent functions of $\sigma^+$ and $\sigma^-$. A nonlinear analogue of this statement in integrable models can be obtained using the so called dressing method (see \cite{Manas} for a review in the context of the principal chiral model). The rough idea is that there exists a pair of gauge transformations \eqref{gauge transf} with parameters $\tilde{g}_{{\rm I}\pm}(x, \sigma^+, \sigma^-)$ which bring the Lax connection into canonical forms depending solely on $\sigma^{\pm}$, respectively:
\begin{equation*}
\tilde{g}_{{\rm I}\pm} J_{\rm I}(x) \tilde{g}_{{\rm I}\pm}^{-1} + (d \tilde{g}_{{\rm I}\pm}) \tilde{g}_{{\rm I}\pm}^{-1} = \frac{j^0_{{\rm I}\pm}(\sigma^{\pm})}{1 \mp x}\, d\sigma^{\pm} \eqqcolon J_{{\rm I}\pm}(x).
\end{equation*}
But moreover, given two such `right and left moving' Lax connections $J_{{\rm I}\pm}(x)$ we can recover the original Lax connection \eqref{Lax connection}. Specifically, we have a pair of maps
\begin{equation*}
\xymatrix{
J_{\rm I}(x, \sigma^+, \sigma^-) \ar @<2pt> @^{->} [rr] ^----{\rm undress} & &
\ar @<2pt> @^{->} [ll] ^----{\rm dress} \big( J_{{\rm I}+}(x, \sigma^+), J_{{\rm I}-}(x, \sigma^-) \big)}
\end{equation*}
referred to as the undressing and dressing transformations. This is effectively the nonlinear counterpart of the (linear) correspondence $X^{\mu}_{\rm I}(\sigma^+, \sigma^-) \rightleftharpoons (X^{\mu}_{{\rm I}+}(\sigma^+), X^{\mu}_{{\rm I}-}(\sigma^-))$ in flat space.

In fact, the analogy with the flat space case goes even further. Suppose we `normalize' the solution $X^{\mu}_{\rm I}(\sigma^+, \sigma^-)$ by requesting that $X^{\mu}_{\rm I}(0, 0) = 0$. This amounts to performing a constant translation on the solution, which is a symmetry of the equations, and the `unnormalized' solution is recovered by adding back the original value of $x^{\mu}_0 \coloneqq X^{\mu}_{\rm I}(0, 0)$. Then the functions $X^{\mu}_{{\rm I}\pm}$ may be defined simply as $X^{\mu}_{{\rm I}+}(\sigma^+) \coloneqq X^{\mu}_{\rm I}(\sigma^+, 0)$ and $X^{\mu}_{{\rm I}-}(\sigma^-) \coloneqq X^{\mu}_{\rm I}(0, \sigma^-)$. Note that they inherit the `normalization' of $X^{\mu}_{\rm I}$ since $X^{\mu}_{{\rm I}\pm}(0) = 0$. The full solution may be obtained from its values on the left and right null rays through the special point $(0, 0) \in W_{\rm I}$,
\begin{equation*}
X^{\mu}_{\rm I}(\sigma^+, \sigma^-) = X^{\mu}_{{\rm I}+}(\sigma^+) + X^{\mu}_{{\rm I}-}(\sigma^-).
\end{equation*}
In particular, the `unnormalized' solution is obtained by adding the constant $x^{\mu}_0$.

The analogous construction in the $\mathbb{R} \times S^3$ case goes as follows. Consider the flat Lax connections $J_{{\rm I}\pm}(x)$ define above but with $j^0_{{\rm I}+}(\sigma^+) \coloneqq j_{{\rm I}+}(\sigma^+, 0)$ and $j^0_{{\rm I}-}(\sigma^-) \coloneqq j_{{\rm I}-}(0, \sigma^-)$. We introduce their local trivializations $\Psi_{{\rm I}\pm}(x, \sigma^+, \sigma^-)$ as in \eqref{aux lin sys} but \emph{normalized} such that $\Psi_{{\rm I}\pm}(x, 0, 0) = {\bf 1}$. Then it turns out that the trivialization $\Psi_{\rm I}(x)$ of the original solution $J_{\rm I}(x)$ normalized by $\Psi_{\rm I}(x, 0, 0) = {\bf 1}$ can be obtained by applying a dressing transformation
\begin{equation*}
\xymatrix{
\big( \Psi_{{\rm I}+}(x), \Psi_{{\rm I}-}(x) \big) \ar [rr] ^----{\rm dress}
& & \Psi_{\rm I}(x).}
\end{equation*}
As in the flat space case, the `unnormalized' extended solution with $\Psi_{\rm I}(x, 0, 0) = \Psi_0$ is obtained by multiplying $\Psi_{\rm I}(x, \sigma^+, \sigma^-)$ on the right by the constant matrix $\Psi_0$, which is a symmetry of the equations \eqref{aux lin sys}.

The purpose of the next subsection is to make these statements precise. We shall use them in the following subsection to obtain a complete description of the outgoing string I.

\subsubsection{Dressing and undressing} \label{sec: (un)dress}

Given initial conditions $j^0_{{\rm I}+}(\sigma^+)$ and $j^0_{{\rm I}-}(\sigma^-)$ on the pair of characteristics $\S_{\pm}$ through $(0, 0)$ we shall reconstruct the full solution $j_{{\rm I}\pm}(\sigma^+, \sigma^-)$.

We start by defining the following pair of flat connections
\begin{subequations} \label{Psi pm def}
\begin{equation} \label{Lax pm}
J_{{\rm I}+}(x) = \frac{j^0_{{\rm I}+}(\sigma^+)}{1 - x} d\sigma^+, \qquad
J_{{\rm I}-}(x) = \frac{j^0_{{\rm I}-}(\sigma^-)}{1 + x} d\sigma^-.
\end{equation}
The corresponding extended solutions normalized at $(0, 0)$ are denoted respectively as $\Psi_{{\rm I}\pm}(x)$, namely
\begin{equation} \label{Psi pm eq}
\big( d\Psi_{{\rm I}\pm}(x) \big) \Psi_{{\rm I}\pm}(x)^{-1} = J_{{\rm I}\pm}(x), \qquad \Psi_{{\rm I}\pm}(x, 0, 0) = {\bf 1}.
\end{equation}
\end{subequations}

\paragraph{Birkhoff factorization.} Consider two small circles $\C^{\pm}$ around $x = \pm 1$ on the Riemann sphere $\mathbb{C}P^1$ and let $I^{\pm}$ be their interiors and $E^{\pm}$ their respective exteriors. We also introduce $\C \coloneqq \C^+ \cup \C^-$, $I \coloneqq I^+ \cup I^-$ and $E \coloneqq E^+ \cap E^-$.
The pair of functions $\Psi_{{\rm I}\pm}(x)$ can be viewed as defining a single function $\C \to SL_2(\mathbb{C})$ and the set of all such smooth maps forms a group $\Loop_{\C} SL_2(\mathbb{C})$ under pointwise matrix multiplication. Consider the Birkhoff factorization problem which consists in writing $\Psi_{{\rm I}\pm}(x)$ as a product of maps in $\Loop_{\C} SL_2(\mathbb{C})$ which extend holomorphically to maps $I \to SL_2(\mathbb{C})$ and $E \to SL_2(\mathbb{C})$, respectively. Specifically,
\begin{equation} \label{Birkhoff dress}
\Psi_{{\rm I}\pm}(x) = \tilde{g}_{\rm I}(x)^{-1} \Psi_{\rm I}(x), \qquad \text{for} \quad x \in \C^{\pm}
\end{equation}
where $\tilde{g}_{\rm I}(x)$ is holomorphic in $I$ and $\Psi_{\rm I}(x)$ is holomorphic in $E$ with $\Psi_{\rm I}(\infty) = {\bf 1}$.

The Birkhoff factorization theorem \cite{PressleySegal} states that there exists an open dense subset of the identity component of $\Loop_{S^1} SL_2(\mathbb{C})$, called the ``big cell'', in which the factorization into loops holomorphic inside and outside the unit circle $S^1 = \{ x \in \mathbb{C} \,|\, |x| = 1 \}$ is possible. In particular, the Birkhoff factorization always exists locally, and this statement remains true also for $\Loop_{\C} SL_2(\mathbb{C})$. Since $\Psi_{{\rm I}\pm}(x, 0, 0) = {\bf 1}$ trivially factorizes into a pair of identity matrices, the existence of a solution to \eqref{Birkhoff dress} is therefore guaranteed for small enough $\sigma^{\pm}$. We shall come back to the question of existence after discussing reality conditions.

If it exists, however, it is easy to see that the factorization \eqref{Birkhoff dress} is unique. For suppose $\Psi_{{\rm I}\pm}(x) = \tilde{g}'_{\rm I}(x)^{-1} \Phi_{\rm I}(x)$ gives another factorization then $\Phi_{\rm I}(x) \Psi_{\rm I}(x)^{-1} = \tilde{g}'_{\rm I}(x) \tilde{g}_{\rm I}(x)^{-1}$, where the left and right hand sides are holomorphic in $E$ and $I$, respectively. However, since they are equal on $\C$, together they define a matrix of holomorphic functions over $\mathbb{C}P^1$ which is therefore constant. But the normalization condition at $\infty \in E$ implies $\Phi_{\rm I}(\infty) \Psi_{\rm I}(\infty)^{-1} = {\bf 1}$ so that this constant is the identity matrix and hence $\tilde{g}'_{\rm I}(x) = \tilde{g}_{\rm I}(x)$ and $\Phi_{\rm I}(x) = \Psi_{\rm I}(x)$.

The Birkhoff factorization \eqref{Birkhoff dress} therefore provides a (local) map $\Psi_{{\rm I}\pm}(x) \mapsto \Psi_{\rm I}(x)$. Before exploiting this map, let us show that it is invertible.
Since the coefficients of the system \eqref{Psi pm def} are holomorphic in $E^{\pm}$, so are its solutions $\Psi_{{\rm I}\pm}(x)$. Furthermore, $\Psi_{{\rm I}\pm}(\infty)$ is a constant matrix which must be the identity by the initial conditions. Therefore given $\Psi_{\rm I}(x)$ one can recover $\Psi_{{\rm I}\pm}(x)$ using the `reverse' Birkhoff factorization problem
\begin{equation} \label{Birkhoff undress}
\Psi_{\rm I}(x) = \tilde{g}_{{\rm I}\pm}(x) \Psi_{{\rm I}\pm}(x), \qquad \text{for} \quad x \in \C^{\pm}
\end{equation}
where $\tilde{g}_{{\rm I}\pm}(x)$ and $\Psi_{{\rm I}\pm}(x)$ are holomorphic in $I^{\pm}$ and $E^{\pm}$, respectively, and with $\Psi_{{\rm I}\pm}(\infty) = {\bf 1}$. This is just a rewriting of \eqref{Birkhoff dress}, where we now consider the matrix $\Psi_{\rm I}(x)$ as given and $\Psi_{{\rm I}\pm}(x)$ as unknowns. In particular, $\tilde{g}_{{\rm I}\pm}(x)$ is the restriction of $\tilde{g}_{\rm I}(x)$ to $I^{\pm}$.

\paragraph{Gauge transformation.} Making use of the second factor in \eqref{Birkhoff dress} we define the following flat connection 1-form,
\begin{equation} \label{Lax reconstruct}
J_{\rm I}(x) \coloneqq \big( d \Psi_{\rm I}(x) \big) \Psi_{\rm I}(x)^{-1}.
\end{equation}
Comparing this with \eqref{Psi pm eq} using the factorization \eqref{Birkhoff undress} we find
\begin{equation} \label{Lax gauge transf pm}
J_{\rm I}(x) = \tilde{g}_{{\rm I}\pm}(x) J_{{\rm I}\pm}(x) \tilde{g}_{{\rm I}\pm}(x)^{-1} + \big( d\tilde{g}_{{\rm I}\pm}(x) \big) \tilde{g}_{{\rm I}\pm}(x)^{-1}.
\end{equation}
This shows that \eqref{Lax reconstruct} is related by a gauge transformation to each of the Lax connections \eqref{Lax pm}, in the sense of \eqref{gauge transf} with parameter $\tilde{g} = \tilde{g}_{{\rm I}\pm}(x)$. As discussed in section \ref{sec: integr}, in order for this gauge transformation to be of any use we must show that it preserves the analytic structure of the Lax connection.

It follows from its definition \eqref{Lax reconstruct} that $J_{\rm I}(x)$ is holomorphic in $E$ and vanishes at $x = \infty$. Its behaviour in $I$ can be deduced from the alternative expressions \eqref{Lax gauge transf pm}. Indeed, the second term in this equation is holomorphic in $I^{\pm}$ whereas the first has a simple pole at $x = \pm 1$ with residue proportional to $d\sigma^{\pm}$. By Mittag-Leffler's theorem this information uniquely specifies $J_{\rm I}(x)$ so we can write
\begin{equation*}
J_{\rm I}(x) = \frac{j_{{\rm I}+}}{1 - x} d \sigma^+ + \frac{j_{{\rm I}-}}{1 + x} d \sigma^-,
\end{equation*}
for some functions $j_{{\rm I}\pm}(\sigma^+, \sigma^-)$. Since $J_{\rm I}(x)$ is flat by definition \eqref{Lax reconstruct}, it follows that $j_{{\rm I}\pm}$ satisfy the equations of the principal chiral model.

\paragraph{Cauchy data.} It remains to show that the initial data of $j_{{\rm I}\pm}$ along the characteristics $\S_{\pm}$ through $(0, 0)$ coincides with $j^0_{{\rm I}+}(\sigma^+)$ and $j^0_{{\rm I}-}(\sigma^-)$. To show this, consider \eqref{Lax reconstruct} in light-cone coordinates,
\begin{equation*}
\big( \partial_+ \Psi_{\rm I}(x) \big) \Psi_{\rm I}(x)^{-1} = \frac{j_{{\rm I}+}}{1 - x}, \qquad \big( \partial_- \Psi_{\rm I}(x) \big) \Psi_{\rm I}(x)^{-1} = \frac{j_{{\rm I}-}}{1 + x}.
\end{equation*}
Setting $\sigma^- = 0$ in the first equation, we see that $\Psi_{\rm I}(x, \sigma^+, 0)$ is holomorphic in $E^+$ since the coefficient of the equation are. But then the solution of the factorization problem \eqref{Birkhoff undress} when $\sigma^- = 0$ is simply given by $\Psi_{{\rm I}+}(x, \sigma^+, 0) = \Psi_{\rm I}(x, \sigma^+, 0)$ and $\tilde{g}_{{\rm I}+}(x, \sigma^+, 0) = {\bf 1}$. Likewise, setting $\sigma^+ = 0$ in the second equation, we find that $\Psi_{\rm I}(x, 0, \sigma^-)$ is holomorphic in $E^-$ which in turn implies $\Psi_{{\rm I}-}(x, 0, \sigma^-) = \Psi_{\rm I}(x, 0, \sigma^-)$. In particular, this yields the desired result
\begin{equation*}
j^0_{{\rm I}+}(\sigma^+) = j_{{\rm I}+}(\sigma^+, 0), \qquad
j^0_{{\rm I}-}(\sigma^-) = j_{{\rm I}-}(\sigma^-, 0).
\end{equation*}

\paragraph{Reality conditions.} Since the circles $\C^{\pm}$ are centered around $x = \pm 1$ they are invariant under conjugation $x \mapsto \bar{x}$. The involution $\hat{\tau}$ therefore sends $\Loop_{\C} SL_2(\mathbb{C})$ to itself and its fixed point subset defines the twisted loop group
\begin{equation*}
\Loop_{\C}^{\hat{\tau}} SL_2(\mathbb{C}) \coloneqq \{ \Psi \in \Loop_{\C} SL_2(\mathbb{C}) \, | \, \hat{\tau}(\Psi) = \Psi \}.
\end{equation*}
It is straightforward to show that the Birkhoff factorization \eqref{Birkhoff dress} restricts to this subgroup. Indeed, suppose $(\Psi_{{\rm I}+}, \Psi_{{\rm I}-}) \in \Loop_{\C}^{\hat{\tau}} SL_2(\mathbb{C})$, then applying $\hat{\tau}$ to \eqref{Birkhoff dress} yields the factorization
\begin{equation*}
\Psi_{{\rm I}\pm}(x) = \tau\big( \tilde{g}_{\rm I}(\bar{x}) \big)^{-1} \tau \big( \Psi_{\rm I}(\bar{x}) \big), \qquad \text{for} \quad x \in \C^{\pm},
\end{equation*}
where $\tau\big( \tilde{g}_{\rm I}(\bar{x}) \big)$ and $\tau \big( \Psi_{\rm I}(\bar{x}) \big)$ are holomorphic in $I$ and $E$, respectively, with $\tau \big( \Psi_{\rm I}(\infty) \big) = {\bf 1}$.
Therefore, by the uniqueness of the Birkhoff factorization \eqref{Birkhoff dress} it follows that $\tilde{g}_{\rm I}$ and $\Psi_{\rm I}$ are also in $\Loop_{\C}^{\hat{\tau}} SL_2(\mathbb{C})$, as claimed.

\paragraph{Existence.} We are finally in a position to address the question of existence of the factorization \eqref{Birkhoff dress}. The reason for postponing this issue until now is that although the Birkhoff factorization in $\Loop_{S^1} SL_2(\mathbb{C})$ is only possible on a dense open subset, it turns out \cite{TUhl, Brander} that for the fixed point subgroup $\Loop^{\hat{\tau}}_{S^1} SL_2(\mathbb{C})$ with respect to a complex anti-linear involution $\hat{\tau}$ of the type \eqref{tau inv}, the Birkhoff factorization \emph{always} exists.
In other words, $\Loop_{S^1}^{\hat{\tau}} SL_2(\mathbb{C})$ is connected and the ``big cell'' in this case is the whole of $\Loop_{S^1}^{\hat{\tau}} SL_2(\mathbb{C})$ so that the Birkhoff decomposition is \emph{global}.

This can be used to prove the desired factorization \eqref{Birkhoff dress} as follows. First of all, consider linear fractional transformations $f^{\pm}$, with real coefficients, mapping $S^1$ to $\C^{\pm}$ and the unit disk $\{ x \in \mathbb{C} \, | \, |x| < 1 \}$ to $I^{\pm}$. This allows us to reduce the Birkhoff facotrization of $\Loop_{\C^{\pm}}^{\hat{\tau}} SL_2(\mathbb{C})$ to that of $\Loop_{S^1}^{\hat{\tau}} SL_2(\mathbb{C})$. In other words, we can decompose any $\Psi_{\pm} \in \Loop_{\C^{\pm}}^{\hat{\tau}} SL_2(\mathbb{C})$ as a product $\Phi^I_{\pm} \Phi^E_{\pm}$ where $\Phi^I_{\pm} \in \Loop_{\C^{\pm}}^{\hat{\tau}} SL_2(\mathbb{C})$ and $\Phi^E_{\pm} \in \Loop_{\C^{\pm}}^{\hat{\tau}} SL_2(\mathbb{C})$ extend holomorphically to $I^{\pm}$ and $E^{\pm}$ respectively, with $\Phi^E_{\pm}(\infty) = {\bf 1}$.

Let $(\Phi_+, \Phi_-) \in \Loop_{\C} SL_2(\mathbb{C})$ denote the loop over $\C = \C^+ \cup \C^-$ defined by the pair of loops $\Phi_{\pm} \in \Loop_{\C^{\pm}} SL_2(\mathbb{C})$ over $\C^{\pm}$. Then the element $(\Psi_{{\rm I}+}, \Psi_{{\rm I}-}) \in \Loop_{\C} SL_2(\mathbb{C})$ can be factorized as
\begin{align*}
(\Psi_{{\rm I}+}, \Psi_{{\rm I}-}) &= (\Psi_{{\rm I}+}, {\bf 1}) ({\bf 1}, \Psi_{{\rm I}-}) = (\Phi^I_+, {\bf 1}) (\Phi^E_+, {\bf 1}) ({\bf 1}, \Psi_{{\rm I}-})\\
&= (\Phi^I_+, {\bf 1}) \big( {\bf 1}, \Psi_{{\rm I}-} (\Phi^E_+)^{-1} \big) (\Phi^E_+, \Phi^E_+) = (\Phi^I_+, {\bf 1}) \big( {\bf 1}, \Phi^I_- \Phi^E_- \big) (\Phi^E_+, \Phi^E_+),\\
&= \big( \Phi^I_+ (\Phi^E_-)^{-1}, \Phi^I_- \big) (\Phi^E_- \Phi^E_+, \Phi^E_- \Phi^E_+),
\end{align*}
where in the first line we have introduced the factorization $\Psi_{{\rm I}+} = \Phi^I_+ \Phi^E_+$ in $\Loop^{\hat{\tau}}_{\C^+} SL_2(\mathbb{C})$, and in the second line the factorization $\Psi_{{\rm I}-} (\Phi^E_+)^{-1} = \Phi^I_- \Phi^E_-$ in $\Loop^{\hat{\tau}}_{\C^-} SL_2(\mathbb{C})$. The last line then gives the desired factorization \eqref{Birkhoff dress} since $\tilde{g}_{{\rm I}+} \coloneqq \Phi^I_+ (\Phi^E_-)^{-1}$, $\tilde{g}_{{\rm I}-} \coloneqq \Phi^I_-$ and $\Psi_{\rm I} \coloneqq \Phi^E_- \Phi^E_+$ are holomorphic in $I^+$, $I^-$ and $E$, respectively, with $\Psi_{\rm I}(\infty) = {\bf 1}$.

\subsubsection{Dressing the outgoing strings}

\begin{figure}[h]
\centering
\begin{tabular}{ccccc}
\psfrag{0}{\gray $0^{\rm th}$}
\psfrag{1}{\gray $1^{\rm st}$}
\psfrag{2}{\gray $2^{\rm nd}$}
\psfrag{s1}{\scriptsize $p_1 \! = \! (0,0)$}
\psfrag{s2}{\scriptsize $p_2$}
\psfrag{s3}{\scriptsize $p_3$}
\raisebox{1mm}{\includegraphics[width=25mm]{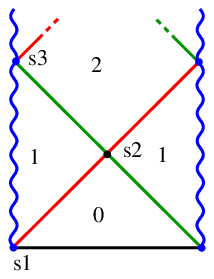}} & $\qquad\qquad\qquad$ &
\psfrag{Sp}{\scriptsize $\S_+$}
\psfrag{Sm}{\scriptsize $\S_-$}
\psfrag{pi}{\scriptsize $p_i$}
\psfrag{pip1}{\scriptsize $p_{i+\!1}$}
\psfrag{i}{\gray $i^{\rm th}$}
\psfrag{ip1}{\scriptsize \gray $(i\!+\!1)^{\rm st}$}
\includegraphics[width=25mm]{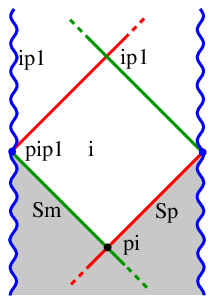} & $\qquad\qquad\qquad$ &
\psfrag{Sp}{\scriptsize $\S_+$}
\psfrag{Sm}{\scriptsize $\S_-$}
\psfrag{pi}{\scriptsize $p_i$}
\psfrag{pip1}{\scriptsize $p_{i+\!1}$}
\psfrag{i}{\gray $i^{\rm th}$}
\psfrag{ip1}{\scriptsize \gray $(i\!+\!1)^{\rm st}$}
\psfrag{dress}{\scriptsize \purple dress}
\includegraphics[width=25mm]{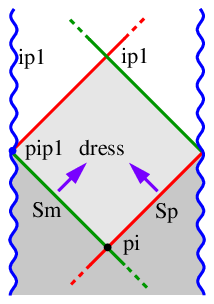}\\
$(a)$ &  & $(b)$ & & $(c)$
\end{tabular}
\caption{$(a)$ We enumerate the different tiles in the tessellation created by the null rays $\S_{\pm}$ emanating from the singularity. In particular, the $0^{\rm th}$ tile contains the Cauchy surface.\\
$(b)$ The shaded region represents the part of the solution already determined. For illustration purposes, we have cut the cylinder in such a way that the $i^{\rm th}$ tile appears whole.\\
$(c)$ The inductive step consists in solving the Cauchy problem on the $i^{\rm th}$ tile, taking for Cauchy data along $\S_{\pm}$ the value of $j^{(i-1)}_{{\rm I}\pm}$ on the boundary of the $(i-1)^{\rm st}$ tile.}
\label{fig: inductive step}
\end{figure}

Putting together the results of this section we obtain a recursive algorithm for constructing the outgoing string solution I, one null-tile at a time. Since the tiles are naturally ordered we label them by integers, the $0^{\rm th}$ tile being the (half) tile introduced in section \ref{sec: abs else S3} and the $i^{\rm th}$ tile ($i \geq 1$) is defined by its lowest point being at (see Figure \ref{fig: inductive step}$(a)$)
\begin{equation*}
p_i \coloneqq (\sigma^+_i, \sigma^-_i) = \left\{ \begin{array}{ll} ( k \, a \pi, (k-1) a \pi ) & \text{for} \;\; i = 2 k,\\
( k \, a \pi, k \, a \pi ) & \text{for} \;\; i = 2 k + 1. \end{array}
\right.
\end{equation*}
The outgoing string can now be constructed recursively as follows.

\paragraph{Initial step.} From section \ref{sec: abs else S3} we know that the solution on the $0^{\rm th}$ tile is simply given by extending the original solution of the incoming string, see Figure \ref{fig: null characteristics}$(b)$.

\paragraph{Inductive step.} Now given the solution on the $(i-1)^{\rm st}$ tile, the solution on the $i^{\rm th}$ tile can be obtained as follows, see Figure \ref{fig: inductive step}$(b)$-$(c)$. Its Cauchy data consists of the values of $j_{{\rm I}\pm}$ along the two null rays $\S_{\pm}$ connecting it to the $(i-1)^{\rm st}$ tile.
Yet by section \ref{sec: prop sing S3} we know that the components $j_{{\rm I}\pm}$ are continuous across the singular null rays $\S_{\pm}$, respectively. Therefore the Cauchy data for the $i^{\rm th}$ tile is completely specified by the solution on the $(i-1)^{\rm st}$ tile as
\begin{equation} \label{ith tile ic}
j^0_{{\rm I}+}(\sigma^+) = j^{(i-1)}_{{\rm I}+}(\sigma^+, \sigma^-_i), \qquad
j^0_{{\rm I}-}(\sigma^-) = j^{(i-1)}_{{\rm I}-}(\sigma^+_i, \sigma^-).
\end{equation}
The solution of the corresponding Cauchy problem is now obtained by applying the dressing transformation of section \ref{sec: (un)dress} with the point $(0, 0)$ there replaced by $(\sigma^+_i, \sigma^-_i)$.

\begin{figure}[h]
\centering
\psfrag{pi}{\scriptsize ${\bf 1}$}
\psfrag{pip}{\scriptsize $\Psi^{(i)}_{{\rm I}+}(x)$}
\psfrag{pim}{\scriptsize $\Psi^{(i)}_{{\rm I}-}(x)$}
\psfrag{pj}{\scriptsize $\Psi^{(i)}_{{\rm I} \, n}(x)$}
\includegraphics[width=35mm]{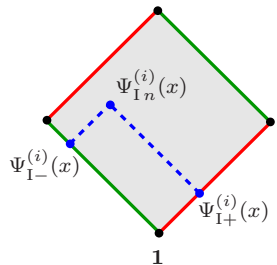}
\caption{The normalized extended solution $\Psi^{(i)}_{{\rm I} \, n}(x, \sigma^+, \sigma^-)$ at the point $(\sigma^+, \sigma^-)$ in the $i^{\rm th}$ tile is obtained from the Birkhoff factorization of the pair $\Psi^{(i)}_{{\rm I}\pm}(x, \sigma^\pm)$ defined at the boundary points $(\sigma^+, \sigma^-_i)$ and $(\sigma^+_i, \sigma^-)$, respectively.}
\label{fig: ith tile}
\end{figure}

The first step requires solving the system \eqref{Psi pm def} with $j^0_{{\rm I}\pm}(\sigma^{\pm})$ given by \eqref{ith tile ic}. However, it is easy to see that the solution can be expressed in terms of the extended solution on the previous $(i-1)^{\rm st}$ tile, since this satisfies the same equations but with a different normalization. Specifically, we have
\begin{subequations} \label{recursive formula}
\begin{equation} \label{recursive formula 1}
\begin{split}
\Psi^{(i)}_{{\rm I}+}(x, \sigma^+) &= \Psi^{(i-1)}_{\rm I}(x, \sigma^+, \sigma^-_i) \Psi^{(i-1)}_{\rm I}(x, \sigma^+_i, \sigma^-_i)^{-1},\\
\Psi^{(i)}_{{\rm I}-}(x, \sigma^-) &= \Psi^{(i-1)}_{\rm I}(x, \sigma^+_i, \sigma^-) \Psi^{(i-1)}_{\rm I}(x, \sigma^+_i, \sigma^-_i)^{-1}.
\end{split}
\end{equation}
Next we perform the Birkhoff factorization \eqref{Birkhoff dress} of $\Psi^{(i)}_{{\rm I}\pm}(x, \sigma^{\pm})$, namely
\begin{equation*}
\Psi^{(i)}_{{\rm I}\pm}(x, \sigma^{\pm}) = \tilde{g}^{(i)}_{\rm I}(x, \sigma^+, \sigma^-)^{-1} \Psi^{(i)}_{{\rm I} \, n}(x, \sigma^+, \sigma^-), \qquad \text{for} \quad x \in \C^{\pm}.
\end{equation*}
The second factor on the right defines the `normalized' extended solution on the $i^{\rm th}$ tile, see Figure \ref{fig: ith tile}. Finally, the `unnormalized' extended solution is now given by
\begin{equation} \label{recursive formula2}
\Psi^{(i)}_{\rm I}(x, \sigma^+, \sigma^-) = \Psi^{(i)}_{{\rm I} \, n}(x, \sigma^+, \sigma^-) \Psi^{(i-1)}_{\rm I}(x, \sigma^+_i, \sigma^-_i).
\end{equation}
\end{subequations}
Equations \eqref{recursive formula} provide the desired recursive formula expressing the extended solution on the $i^{\rm th}$ tile in terms of that on the $(i-1)^{\rm st}$ tile through the use of a Birkhoff factorization.

\section{Conclusions and Outlook}

The classical integrability of the superstring $\sigma$-model on $AdS_5 \times S^5$ has so far played a vital role in the classification of its finite-gap solutions \cite{Classification} as well as their reconstruction \cite{Reconstruction} in the subsector $\mathbb{R} \times S^3$. In this article we made a first step beyond solutions with cylindrical worldsheet by constructing the general splitting solution in $\mathbb{R} \times S^3$. Although the worldsheets of these new solutions have the topology of a pair of pants, the integrability of the $\sigma$-model also played an essential role in their construction. This is no surprise since after all the Lax connection is a local object on the worldsheet.

Specifically, given any string solution with cylindrical worldsheet on $\mathbb{R} \times S^3$, which satisfies the self-intersection property at some instant in time, we constructed the pair of outgoing strings resulting from the split. This was achieved by reducing the problem to factorization in a loop group, as is usual in classical integrable systems. It would be important to investigate further the possibility of solving these Birkhoff factorization problems more explicitly, for instance in terms of Riemann $\theta$-functions.

An example of initial string could be a finite-gap string, the moduli of which are encoded in a finite-genus algebraic curve.
In fact, since the outgoing strings are uniquely determined by their Cauchy data which in turn is given by the incoming string, the entire splitting solution is uniquely characterized by the same algebraic curve as the initial string. The difference between these two solutions will show up in the behaviour of the angle variables, encoded in the algebro-geometric language as a divisor on the curve \cite{Dorey:2006mx}, at the moment of the splitting.
It would be interesting, though, to have a more algebraic characterization of the self-intersection property at the level of the curve and the divisor.

This brings up the curious observation that for a given initial string there can more than one possible evolution, depending on whether or not we choose the string to split at $\tau = 0$. This existence of multiple different solutions for the same set of Cauchy data at $\tau < 0$ is merely a consequence of the fact that the topology of the worldsheet is not determined by the dynamics but rather fixed by hand from the outset. Another way to phrase this is to note that since the metric is not dynamical, it must be fixed prior to solving the equations. Its conformal class then reflects the underlying topology of the worldsheet. For instance, on the cylinder the metric can be made globally flat, whereas on the pair of pants it must be degenerate at the singular point.


Throughout our construction we have assumed the initial string to be smooth at $\tau = 0$. Since the pair of outgoing strings are not smooth along the null rays through the splitting point, it is therefore not immediate how to describe their potential further splitting. This would first require a slight generalization of the construction to include initial solutions with a discrete set of singularities propagating along null lines.
It would also be interesting to study the joining of two classical strings in a similar fashion, as well as classical solutions exhibiting more general worldsheet topology. This is a novel possibility on curved backgrounds such as $\mathbb{R} \times S^3$ since the products of a split can eventually meet again and join.

We emphasize that splitting strings are solutions of an initial value problem for a system of hyperbolic (Lorentzian) differential equations. Such solutions, which describe a complicated splitting process in $S^3 \subset S^5$, should therefore be relevant for the semiclassical computation of 3-point functions in the \emph{Minkowskian} approach of \cite{Janik:2010gc}. By comparison, the problem of constructing minimal surfaces in \emph{Euclidean} AdS ending at certain points on the boundary is a very different one. It can \emph{a priori} be phrased as a boundary value problem for a system of elliptic (Euclidean) differential equations. However, the classical minimal surface dominating a 3-point correlation function at strong coupling should also contain extra information, on each leg, about the type of operator inserted at the boundary.

Another promising approach for computing strong coupling 3-point functions directly within the Euclidean formalism is the vertex operator approach \cite{Tseytlin:2003ac}. The main obstacle in this direction is the construction of vertex operators corresponding to finite-gap solutions. The insertion of three such operators in the path integral should produce the correct boundary conditions for the minimal surface mentioned above. Because finite-gap solutions have Lorentzian worldsheets while minimal surfaces have Euclidean signature, one would naively expect the vertex operator to create a Euclidean continuation of the finite-gap solution. Such a relation is currently only understood for 2-point functions \cite{Buchbinder:2010vw}.



Finally, our construction should have a natural generalization to $AdS_5 \times S^5$ superstrings or more generally to $\mathbb{Z}_4$-graded supercoset $\sigma$-models \cite{Zarembo:2010sg}. Indeed, the loop group factorization discussed here is the global counterpart of the loop algebra decomposition discussed in \cite{Vicedo:2010qd}.

\section*{Acknowledgements}

This work was motivated by a talk of R. Janik presented at the workshop ``From Sigma Models to Four-dimensional QFT'' at DESY, Hamburg. I would like to thank Y. Aisaka, T. W. Brown, N. Dorey, R. Janik, M. C. Reed, J. Teschner, K. P. Tod, A. A. Tseytlin and K. Zarembo for interesting discussions.

\appendix

\end{document}